**Time-resolved magneto-photoluminescence studies of magnetic polaron dynamics in type-II quantum dots**

B. Barman[1], R. Oszwałdowski[1,6], L. Schweidenback[1], A. H. Russ[1], J. M. Pientka[1,2], Y. Tsai[1,] W-C. Chou[3], W. C. Fan[3], J. R. Murphy[1], A. N. Cartwright[4], I. R. Sellers[5], A. G. Petukhov[6], I. Žutić[1], B. D. McCombe[1] and A. Petrou[1]

[1] *Department of Physics, University at Buffalo SUNY, Buffalo, NY 14260 USA*

[2] *Department of Physics, St. Bonaventure University, St. Bonaventure, NY 14778 USA*

[3] *Department of Electro-physics, National Chiao Tung University, Hsinchu 300, Taiwan*

[4] *Department of Electrical & Electronic Engineering, University at Buffalo SUNY, Buffalo, NY 14260 USA*

[5] *Department of Physics and Astronomy, University of Oklahoma, Norman, OK 73019 USA*

[6] *Department of Physics, South Dakota School of Mines & Technology, Rapid City, SD 57701 USA*

We used continuous wave photoluminescence (cw-PL) and time resolved photoluminescence (TR-PL) spectroscopy to compare the properties of magnetic polarons (MP) in two related spatially indirect II-VI epitaxially grown quantum dot systems. In the ZnTe/(Zn,Mn)Se system the holes are confined in the non-magnetic ZnTe quantum dots (QDs), and the electrons reside in the magnetic (Zn,Mn)Se matrix. On the other hand, in the (Zn,Mn)Te/ZnSe system, the holes are confined in the magnetic (Zn,Mn)Te QDs, while the electrons remain in the surrounding non-magnetic ZnSe matrix. The magnetic polaron formation energies $E_{MP}$ in both systems were measured from the temporal red-shift of the band-edge emission. The magnetic polaron exhibits distinct characteristics depending on the location of the Mn ions. In the ZnTe/(Zn,Mn)Se system the magnetic polaron shows conventional behavior with $E_{MP}$ decreasing with increasing temperature *T* and increasing magnetic field *B*. In contrast, $E_{MP}$ in the (Zn,Mn)Te/ZnSe system has unconventional dependence on temperature *T* and magnetic field *B*; $E_{MP}$ is weakly dependent on $T$ as well as on $B$. We discuss a possible origin for such a striking difference in the MP properties in two closely related QD systems.

**I. INTRODUCTION**

Quantum dots (QDs), also known as artificial atoms, can allow versatile control of the number of carriers, their spin, Coulomb interactions, and quantum confinement.[1-4] Compared to their bulk counterparts,[5-9] magnetically doped semiconductor QDs could provide control of the magnetic ordering, [10-16] with the onset of magnetization at substantially higher temperatures.[17-21] Experiments typically focus on Mn-doped II-VI and III-V QDs, in which it is possible to include both single[22-25] and several magnetic impurities, [17-21, 26-40] having similarities with nuclear spins.[41, 42] In the first case (single magnetic ion), such systems could be considered as potential quantum bits, quantum memories, or probes to detect an unconventional orbital ordering.[17, 23-25, 43] In the second case, the presence of several magnetic ions can lead to the formation of a magnetic polaron (MP), a long-standing research topic in magnetic semiconductors.[5-8, 44] MP can be viewed as a cluster of localized magnetic ion spins, aligned through an exchange interaction with the spin of a confined carrier. Initial studies in bulk systems involved MPs with a carrier spin bound to an impurity center (donor or acceptor).[8] In contrast, in semiconductor nanostructures with reduced dimensionality, the confinement removes the need for the presence of impurities and enhances the stability of the MPs.[7] As depicted in Fig. 1, after a sufficiently long time interval after photoexcitation (comparable to MP formation time), Mn-spins in II-VI systems typically couple ferromagnetically with electron spins and antiferromagnetically with hole spins. The simultaneous presence of carriers and Mn ions in QDs result in the formation of MP through lowering of the exciton energy by an amount as shown in

Fig. 1. Two main classes of magnetic QDs have been investigated; those grown using molecular beam epitaxy (MBE)[7, 17, 20-25, 29-35, 39] and those that are solution-processed, known as colloidal QDs.[12, 26-28, 36, 37, 40] Despite entirely different growth procedures, in both classes of QDs the MP formation is associated with the observed magnetic ordering.[12, 20-22, 30, 37] Several interesting effects have been attributed to MPs in nanostructures, such as the long "spin memory" times in (Cd,Mn)Te QDs[30], giant magnetoresistance in ErAs:GaAs nanocomposites,[45] and room-temperature ferromagnetic ordering in MnGe QDs.[21] The temporal evolution of the MP (Fig. 1) can be studied using time-resolved photoluminescence spectroscopy (TR-PL).[39] In these experiments large (tens of meVs) red shifts of the photoluminescence (PL) peak energy are observed as a function of time delay between laser excitation and PL detection.

The majority of published work describe studies of magnetic QDs with type-I band alignment,[7, 39] where the location of electrons and holes coincide spatially. In this work, we investigate TR-PL measurements in QD structures with type-II band alignment where the holes are confined in the QDs while the electrons reside in the surrounding matrix. Schematic diagrams of type-I and type-II alignment are shown in Fig. 2(a) and Fig. 2(b) respectively. We have studied two closely related ZnTe/ZnSe QD systems, grown using the same MBE process. In sample 1, $Mn^{2+}$ ions are incorporated in the matrix: ZnTe/(Zn,Mn)Se, while in sample 3, $Mn^{2+}$ ions are in the QDs region: (Zn,Mn)Te/ZnSe.

These type-II structures offer two clear advantages for the study of MP dynamics over type-I QDs: (i) In type-II QDs, the photoexcited electrons and holes are spatially separated, i.e., have a small wavefunction overlap. As a result, the recombination time is comparable to or exceeds the MP formation time, and thus the MP has sufficient time to develop before electron-hole recombination takes place. (ii) The energy of the spatially-indirect interband transitions in (Zn,Mn)Te/ZnSe and ZnTe/(Zn,Mn)Se QDs is 1.9 eV, i.e., it lies below the competing Mn internal transition at 2.2 eV.[7] Thus, most of the luminescence intensity appears in the interband recombination channel.

Our experimental results show that the MP in the closely related ZnTe/(Zn,Mn)Se and (Zn,Mn)Te/ZnSe QD structures show a strikingly different dependence on temperature and magnetic field. This difference in MP properties in the two systems is attributed to the difference in the location and magnetic susceptibility of the Mn ions.

## II. EXPERIMENTAL

We have used 4 samples in this study. Sample 1 is a ZnTe/(Zn,Mn)Se QD structure while sample 3 is a (Zn,Mn)Te/ZnSe QD structure. Samples 2 and 4 are the non-magnetic references for samples 1 and 3 respectively, grown in the same growth runs as their magnetic counterparts. In samples 1 and 3, the average Mn composition is 5.2% as measured by Energy Dispersive X-Ray Spectroscopy. All samples were grown by MBE on (100) GaAs substrates. The details of sample growth have been given elsewhere[33]. All the QDs have a disk shape with an average 20nm base diameter and a height of 3nm as determined from cross sectional transmission electron microscopy studies. In Fig. 3(a) we show a schematic of the disk-shaped QDs. The samples have been placed in a variable temperature optical magnet cryostat operating in the 5-140 K temperature range. The magnetic field was applied along the direction perpendicular to the QD layers, defined as the *z*-axis, as is shown in Fig. 3(a). The Faraday geometry was used, with the emitted light propagating along the magnetic field. The possible interband recombination channels in the Faraday geometry are illustrated in Fig. 3(b): Spin-down electrons ($m_S = -1/2$) recombine with spin-up holes ($m_J = +3/2$) emitting $\sigma_+$ photons; spin-up electrons ($m_S = +1/2$) recombine with spin down holes ($m_J = -3/2$) emitting $\sigma_-$ photons. In Fig. 3(b), we have included only the heavy holes. The strong valence-band confinement drives the light-hole levels well below the heavy-hole levels. As a result, the light holes do not contribute to the PL spectra.[46] A combination of quarter-wave plate and linear polarizer was placed in appropriate configurations before the spectrometer entrance slit to separate the $\sigma_+$ from the $\sigma_-$ components of the emission. The continuous wave (cw) PL was excited by the linearly polarized 488 nm (2.54 eV) line of an argon-ion laser. The cw PL was analyzed by a single monochromator equipped with a charged coupled device (CCD) multichannel detector. The TR-PL was excited at 400 nm by a frequency doubled pulsed laser system (repetition rate = 250 kHz, pulse duration ~200 fs). The TR-PL was spectrally resolved by a monochromator and temporally analyzed by a streak camera having a temporal resolution of 40 ps. The TR-PL data were divided into time slices. Each slice was fitted with a Gaussian that yielded the peak position and the intensity as a function of time delay $\Delta t$.

## III. RESULTS

We first discuss the cw PL results. In Fig. 4 we plot the peak position energies of the PL from sample 1 (triangles) and sample 3 (circles) as function of applied magnetic field *B*. Sample 1 has a red shift of 12 meV at *B* = 6 tesla. In this sample the emission is due to recombination of electrons in the magnetic (Zn,Mn)Se matrix with holes confined in the ZnTe QDs. Therefore the red shift is mainly due to the exchange interaction of electron spins with the spins of the Mn ions in the matrix. An additional contribution to the red shift due to the interaction of holes in the ZnTe QDs with Mn that diffused into the non-magnetic QDs cannot be excluded.[47] The red shift of PL in sample 1 decreases with increasing temperature. These data strongly indicate that the (Zn,Mn)Se matrix in sample 1 is in the paramagnetic phase with large, temperature-sensitive, band Zeeman splittings. In contrast, sample 3 exhibits a smaller red shift of 4 meV at $B = 6$ tesla even though the holes are confined in the magnetic (Zn,Mn)Te quantum dots. The unusual result for sample 3 shown in Fig. 4 will be discussed in Section IV.

We next turn to the time-resolved measurements. In Fig. 5(a) and 5(b), we show the Gaussian fits derived from the TR-PL spectra in cascade form from sample 1 and sample 3, respectively recorded at $B =0$ and $T = 7$K. The exciting laser pulse arrived at $t = 1.96$ ns for both spectra. Photon collection was chosen to always start earlier than the pulse arrival to ensure that the entire PL time evolution was recorded. The time delay $\Delta t$ in the remainder of the text is defined as the difference between the detection time and the pulse arrival time (i.e. difference between the detection time and the photon-collection start time reduced by 1.96 ns in Fig. 5). Immediately after the pulse arrival, the PL peak in Fig. 5(a) and 5(b) is at 1.96eV. The peak energies red shift with increasing time delay reaching a value of 1.91 eV for sample 1 and 1.89 eV for sample 3, at $\Delta t \approx 18$ ns.

Additional information about the time-resolved measurements is shown in Fig. 6 where we plot the peak energies for our samples as function of time delay $\Delta t$. The magnetic samples 1 and 3 show large red shifts (tens of meV) with $\Delta t$ as shown in the lower panels of Fig. 6. For the laser powers used in our experiments the peak energy of the non-magnetic samples 2 and 4 exhibit a smaller but sizeable red shift as shown in the upper panels of Fig. 6(a) and Fig. 6(b). The data from samples 2 and 4 were fitted by a single exponential time evolution with time constants $\tau_2 = 16$ ns and $\tau_4 = 6.9$ ns, respectively. In order to obtain fits for the magnetic samples 1 and 3 [solid yellow lines in Fig. 6(a) and 6(b)], we had to use two decaying exponentials with two corresponding time constants: a fast time constant $\tau_{1F}$ ($\tau_{3F}$) for sample 1 (sample 3) and a slow time constant $\tau_{1S}$ ($\tau_{3S}$). Times $\tau_{1S}$ and $\tau_{3S}$ are comparable to $\tau_2$ and $\tau_4$ respectively. The fitted time constants for all samples are summarized in table 1.

Table 1: Zero field TR-PL parameters

| Sample | $\tau_F$ (ns) | $\tau_S$ (ns) | $\tau$ (ns) | $E_{MP}$ (meV) |
|---|---|---|---|---|
| 1 | 0.35 | 17.2 | | 25.4 |
| 2 | | | 16 | |
| 3 | 0.48 | 7 | | 34.4 |
| 4 | | | 6.9 | |

Given the similarity of $\tau_{1S}$ with $\tau_2$ on one hand, and the similarity of $\tau_{3S}$ with $\tau_4$ on the other, we attribute the entire red shift of the peak position in the non-magnetic samples 2 and 4, and the slower component of the red shift in the magnetic samples 1 and 3 to the same spin-independent mechanism. A possible mechanism could be electric-dipole layer formation at the wetting layer/ZnSe matrix interface. Such dipole layers have been predicted and studied by other groups in ZnSeTe multilayers and type-II GaSb/GaAs quantum wells.[48,49] The total temporal red shifts $R_1$ and $R_3$ of the TR-PL for samples 1 and 3 was determined from the sum of the two energy parameters in the bi-exponential fit of the peak position energy described above. For the non-magnetic samples 2 and 4 there was only one component contributing to the temporal red shifts $R_2$ and $R_4$.

Even though the red shifts, of samples 1 and 2 on one hand and samples 3 and 4 on the other, depend strongly on laser power, the differences $R_1 - R_2$ and $R_3 - R_4$ remain constant over a wide range of laser powers, indicating that $R_1 - R_2$ and $R_3 - R_4$ is not related to the spin-independent mechanism responsible for the slow red shift. Therefore we identify the faster components $\tau_{1F}$ and $\tau_{3F}$ of the red shifts in samples 1 and 3 as the MP formation times ($\tau_{1MP} = \tau_{1F}$ and $\tau_{3MP} = \tau_{3F}$). In a similar fashion we identify the energy differences $R_1 - R_2$ and $R_3 - R_4$ as the MP formation energies, $E_{1MP}$ and $E_{3MP}$ respectively, in the magnetic samples.

**IV. DISCUSSION**

In Fig. 7 (a) [Fig. 7 (b)], we show a schematic of the MP formation for sample 1 (sample 3) for the spin up holes and spin down electrons ($\sigma_+$ polarization recombination channel). The picture with all spins reversed (Mn, electrons, and holes) would correspond to the $\sigma_-$ polarization. Top panels correspond to the picture before electron-hole photo-excitation. The Mn spins, indicated by the orange arrows, are randomly oriented. Middle panels describe the system immediately after photo-excitation and before the MP had time to form. Thus the Mn spins continue to be randomly oriented. The electron spins in both samples are also randomly oriented; in contrast the direction of the hole spins in our flat disc-shaped QDs (height much smaller than the diameter) is determined by the strong confinement and by the spin-orbit interaction to be oriented either parallel or antiparallel with the QD axis (z-axis).[32, 50] Bottom panels show the spin orientation for the Mn, the electron and the hole spins after the MP formation. In both samples the direction of the hole spins defines the orientation of the Mn ion spins which are oriented antiferromagnetically with the hole spins.[6] The electron spins orient themselves ferromagnetically with the Mn spins. The MP formation results in the reduction in energy of the Mn-hole-electron complex by an amount $E_{MP}$ as shown in Fig. 1(c).

The hole-Mn spin exchange interaction is stronger in sample 3 due to the fact that Mn and holes occupy the same space [the (Zn,Mn)Te QDs]. In sample 1 the hole-Mn interaction is weaker and is present due to the penetration of the hole wavefunction in the (Zn,Mn)Se matrix and possible diffusion of Mn in the ZnTe QDs. Therefore it is expected that $E_{1MP} < E_{3MP}$; this is indeed the case as can be seen from the $E_{MP}$ values listed in Table 1. It is clear that the ratio $E_{3MP}/E_{1MP}$ is not equal to the ratio of the exchange constants for holes and electrons, $\beta/\alpha$.[47, 51] This indicates that in sample 1 we may have some diffusion of Mn from the (Zn,Mn)Se matrix into the ZnTe QDs.

The MP energies, $E_{MP}$, for zero magnetic field are plotted as function of temperature in Fig. 8(a) and Fig. 8(b) for samples 1 and 3, respectively. $E_{MP}$ of sample 1 shows the typical temperature dependence, i.e., it decreases with increasing temperature.[39] Surprisingly, $E_{MP}$ of sample 3 has a weak temperature dependence.

The dependences of $E_{MP}$ on magnetic field $B$, at constant temperature, differ between samples 1 and 3 as well, see Fig. 9. Sample 1 [Fig. 9(a)] exhibits the conventional trend, i.e., $E_{MP}$ decreases with increasing $B$.[52] In contrast, $E_{MP}$ in sample 3 is roughly independent of $B$. Weak $B$-field dependence has been reported in CdTe/(Cd,Mn)Te superlattices[53] and also presented in Fig. 7-16 of Ref. 7.

An important model of magnetic polaron formation in DMS was used to successfully analyze spin-flip data of donor bound electrons in (Cd,Mn)Se,[44, 54]

$$E_{MP} = m_0^{-1}\left(J_{ex}/2gm_B N_0\right)^2 h\left(E_{MP}/k_B T\right) W_{eff}^{-1} C(T), \qquad (1)$$

where $J_{ex}$ is the exchange integral for carriers, $N_0$ is the cation density, $g = 2$, $\mu_B$ is the Bohr magneton, and $\Omega_{eff}$ is the effective MP volume. The term $\eta(E_{MP}/k_B T) = \tanh(E_{MP}/k_B T)$.[20] As can be seen from Eq. (1) the magnetic susceptibility $\chi$ has a strong influence on the properties of MPs in (Cd,Mn)Se and DMS systems in general.[44] The use of Eq. (1) is appropriate only if the molecular field $B_m$ is small so that it does not saturate the Mn spins. This is the case for sample 1 which incorporates a (Zn,Mn)Se matrix, characterized by conventional paramagnetic susceptibility that decreases quickly with increasing temperature. From the typical strong temperature dependence (~1/T) of $\chi$, the trend in Fig. 8(a) is consistent with Eq. (1). At the same time, the conventional values of $\chi$ at low temperatures are sufficiently high to allow for significant alignment of the Mn spins in the presence of an applied magnetic field of a few tesla.[6] Due to this alignment, the temporal red-shift of PL, for $B \neq 0$, identified as the MP formation energy $E_{MP}$, is smaller than for $B = 0$. This is because recombination events at zero delay time occur for carrier energies defined by Mn spins partially aligned by $B$, while events at long delay times occur for full Mn spin alignment, as they did for $B=0$. Therefore, the energy difference between the zero-delay and long-delay recombination events (i.e., $E_{MP}$) must be smaller than the same difference at $B = 0$, in agreement with the results of Fig. 9(a).

Turning to sample 3 we see significant differences in $E_{MP}$ (T, B) as compared to sample 1, as well as what would be expected for the MP energy from Eq. (1). Our theoretical description of sample 3 should then reconcile: (a) small red shift with $B$ of the cw PL peak energy, (b) weak dependence of $E_{MP}$ on $T$ and (c) weak dependence of $E_{MP}$ on $B$. In an earlier work (Ref. 20), which included the results for $E_{MP}(T)$ from Fig. 8(b), but neither $E_{MP}(B)$, nor the measurements on samples 1 and 2, an attempt was made to explain aspects (a) and (b) using Eq.

(1) and the assumption of antiferromagnetic coupling of the Mn spins which would give a weak dependence of $\chi$ on $T$. However, additional measurements in the present work suggest a different and a more plausible explanation realizing that from (b) and (c) we should expect that this robust MP behavior is a consequence of a large molecular (exchange) field. Following this motivation, we use Eq. (2) from the paper by Dietl et al.[55] and Eq. (7.3) in Ref. 7, to calculate the molecular field $B_m$ and its mean-field approximation value.

$$B_m = \frac{1}{3\mu_B g} \beta J \left|\psi\left(\vec{r}\right)\right|^2 \approx \frac{1}{3\mu_B g} \beta J \frac{1}{\Omega_{eff}}, \qquad (2)$$

where $\beta$ is the exchange constant for holes in (Cd,Mn)Te, $J = 3/2$ is the hole spin and $\psi$ is the hole wavefunction. Since the hole localization diameter can be smaller than the QD diameter due to alloy and spin disorder scattering,[7,56,57] we treat the effective QD diameter $d'$ and effective QD height $h'$ as adjustable parameters for the $\Omega_{eff}$, given by the expression,

$$\Omega_{eff} = \pi\hbar \frac{h'd'}{3} \sqrt{\frac{2}{m^* \Delta E_{VB}}} \qquad (3)$$

where, $m^* = 0.19 m_e$ is the hole effective mass, and $\Delta E_{VB} = 1\text{eV}$ is the valence band offset. The calculated values of $B_m$ are plotted in Fig. 10 as a function of $d'$ for three values of $h'$. It is clear that $B_m$ increases with decreasing values of $d'$ and $h'$. In order to obtain the high values of $B_m$ suggested by the data from sample 3 shown in Fig. 8(b) and Fig. 9(b), we chose $d'$ to be equal to 5 nm and $h' = h = 3$ nm. This gives a value of $B_m \approx 20$ tesla. The high value of $B_m$ would also explain the small red shift in sample 3 due to the application of an external magnetic field shown in Fig.4. Since sample 3 is a type-II heterostructure with an exciton lifetime longer than the MP formation time, the hole has enough time to polarize the surrounding Mn spins. This results in a relatively small red shift induced by the externally applied magnetic field.

The free energy functional of the magnetic polaron can be expressed as,[44]

$$F_{MP} = \frac{\hbar^2}{2m^* L_{MP}^2} + \frac{\hbar^2 \pi^2}{2m^* h^2} + \frac{m^* \omega^2}{2} L_{MP}^2 - k_B T \ln 2 - k_B T \sum_{j=1}^{N_{Mn}} \ln \left[ \frac{\sinh\left(\frac{2S+1}{2}\left(\beta \rho_{MP}(R_j)/3k_B T\right)\right)}{\sinh\left(\frac{1}{2}\left(\beta \rho_{MP}(R_j)/3k_B T\right)\right)} \right], \quad (4)$$

where $L_{MP}$ is a free parameter that describes the lateral extent of the MP wavefunction $\psi_{MP}$, $\rho_{MP} = (3/2)|\psi_{MP}(\vec{r})|^2$ is the hole spin density, and $\omega$ is the oscillator frequency that describes the lateral confinement of the holes. In Eq. (4) the first two terms represent the kinetic energy, the third term is the potential energy, and the fourth term comes from hole degeneracy. The final term is the exchange energy between hole and Mn spins. For the calculation of the average MP energy, because of the initial localization of the hole, it is sufficient to consider only the last term in Eq. (4). The average MP energy can be obtained from,

$$E_{MP} = F_{MP} - T \frac{\partial F_{MP}}{\partial T}. \quad (5)$$

In Fig. 11 we plot $E_{MP}(T)$ for $d' = 20, 10, 5,$ and $3$ nm. The effective MP temperature $T_{MP}$ can be higher than the lattice temperature of 7 K. The elevated $T_{MP}$ could be due to the high-peak power of the pulsed laser used to excite the TR-PL spectra.[58] In Fig. 11, if we consider $T > 7$ K, we have a weak dependence of $E_{MP}(T)$, close to the results of Fig. 8(b).

In order to calculate $E_{MP}(B)$, we rewrite Eq. (2) as,

$$\frac{\beta \rho_{MP}(\vec{r})}{3} = \frac{\beta J |\psi(\vec{r})|^2}{3} = g\mu_B B_m(\vec{r}). \tag{6}$$

$\beta \rho_{MP}(\vec{r})$ in terms of $B_m(\vec{r})$ and substitute $B_m(\vec{r})$ with $B_m(\vec{r}) + B$. Using Eq. (5), we calculate the average MP energy and subtract the Zeeman shift[59] to obtain $E_{MP}$. In Fig. 12 we plot $E_{MP}(B)$ at T= 7 K, 14 K and 25 K. If we assume that $T_{MP}$ is higher than the lattice temperature of 7 K, we observe: (i) a weak dependence of $E_{MP}(B)$ in agreement with Fig. 9(b); and (ii) $E_{MP}$ for 0<B<4 tesla is close to the experimental value of 35 meV. This model, compared to the preliminary one discussed in 2010 PRB[20], gives a more complete (improved) description of our experiment data on sample 3.

## V. CONCLUSIONS

We have performed magneto-optical studies of magnetic polaron formation in two closely related type-II (spatially indirect) QD systems: ZnTe/(Zn,Mn)Se (sample 1) and (Zn,Mn)Te/ZnSe (sample 3). These were grown by the same experimental group using the same procedures. Likewise, the optical measurements and the corresponding data analysis were also performed in the same way. MP formation was observed in both systems; nevertheless, there are striking differences in their properties. In sample 1 where the magnetic ions are located outside of the QDs in the surrounding matrix, the MP formation energy has a strong temperature and magnetic field dependence, similar to previously studied type-I QDs.[39]

Electrons which are mostly responsible for the MP formation in sample 1 are not strongly localized in the magnetic matrix. Therefore, we would expect that properties of such samples would resemble those of extensively studied bulk systems. Indeed, the MP formation picture developed for donors in bulk DMS provides a very good description for MP properties in sample 1. In contrast, in sample 3 we expect more pronounced quantum confinement effects where the magnetic ordering would arise from exchange interaction of Mn ion spins with the spin of holes that are strongly localized within the QDs.

In order to understand the MP properties in sample 3 we performed calculations of the molecular magnetic field $B_m$, as well as the dependence of the MP energy $E_{MP}$ on $T$ and $B$. If we assume strong hole localization due to alloy and spin disorder scattering,[55] our calculations suggest a large molecular field $B_m$ (>20 tesla). If we make the additional assumption that the pulsed laser excitation resulted in an increase of the hole-Mn system effective temperature above the lattice temperature, our calculations describe adequately the behavior of sample 3 as shown in Fig. 8(b) and Fig. 9(b).

Additional guidance for a suitable theoretical description would come with the availability of new materials systems. Just as in other prior DMS work, we anticipate a transition from bulk-like systems to structures of reduced dimensionality.[7] One important example would be to realize QDs from novel Mn-doped II-II-V DMS which can have independent charge and spin doping and would therefore be suitable to test the MP formation for a wide range of parameters.[60,61,62]

---

ACKNOWLEDGMENTS

This work was supported by U.S. DOE, Office of Science BES, under Award DE-SC0004890 (I.Z., J.M.P.,A.G.P), NSF DMR-1305770 and U.S. ONR N000141310754 (R.O.).

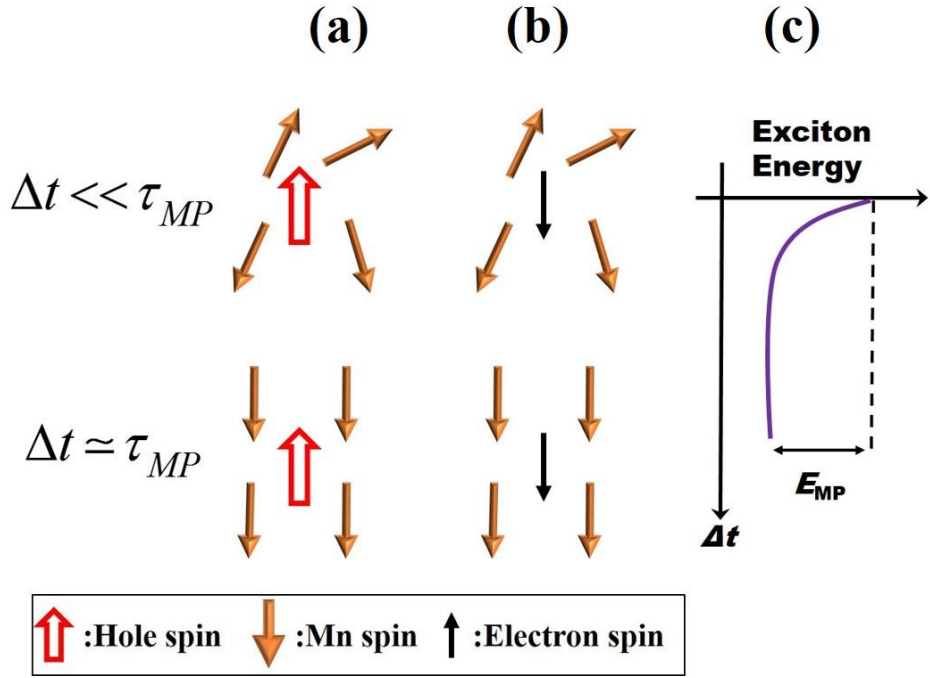

Fig. 1 [(a) and (b)] (color online): Schematic diagram of the formation of magnetic polarons (MP). The red (black) arrows indicate the hole (electron) spin; orange arrows are used for the magnetic ion spins. The hole (electron) spin is antiferromagnetically (ferromagnetically) aligned with the surrounding magnetic ion spins. Here $\Delta t$ is the difference between the PL detection time and the pulse arrival time and $\tau_{MP}$ is the polaron formation time. The upper panels in Fig. 1(a) and Fig. 1(b) depict the situation at early times ($\Delta t \ll \tau_{MP}$) following photoexcitation and before MP is formed. The lower panels in Fig. 1(a) and Fig. 1(b) refer to later times ($\Delta t \cong \tau_{MP}$) with the MP fully formed. (c) A schematic plot of the exciton energy as function of $\Delta t$.

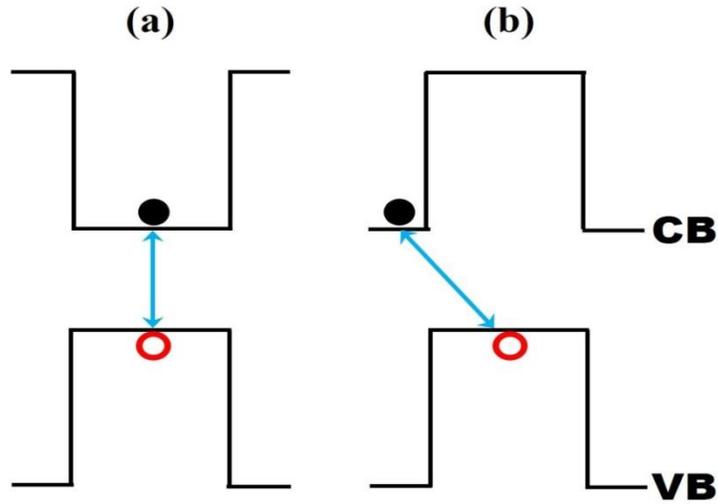

Fig. 2 (color online): Schematic of the band diagram in: (a) Spatially direct (type-I) quantum dots and (b) Spatially indirect (type-II) quantum dots. Here CB and VB indicate the conduction and valence band edges respectively. Electrons (holes) are denoted by full circles (open circles).

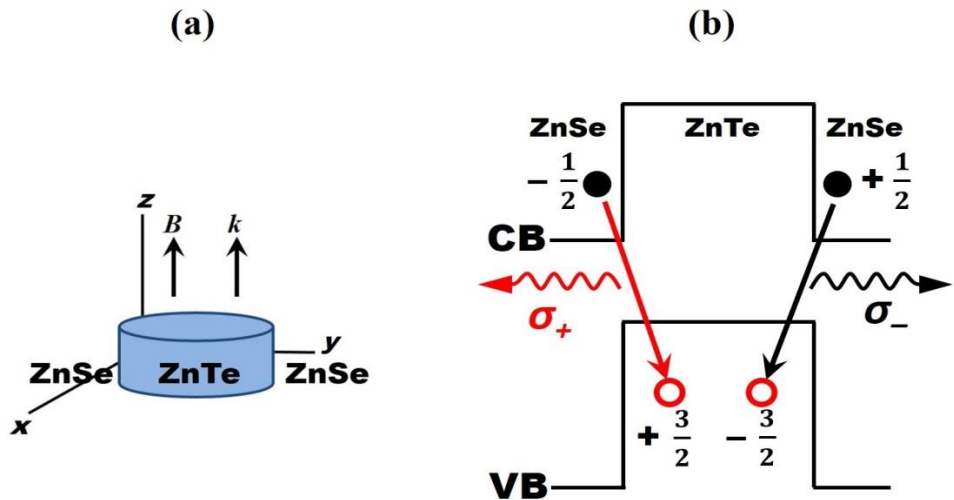

Fig. 3 (color online): (a) Schematic of the disc shaped ZnTe/(Zn,Mn)Se and (Zn,Mn)Te/ZnSe quantum dots used in this work in which the dot diameter is much larger than the height. Here B is the externally applied magnetic field and k is the wave vector of the emitted luminescence (Faraday geometry). (b) Allowed interband radiative transitions in the Faraday geometry.

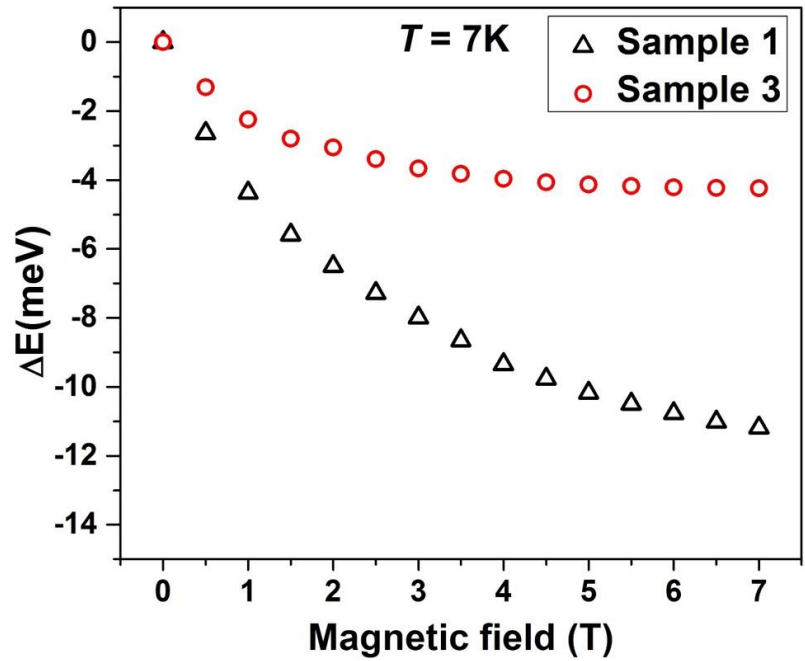

Fig. 4 (color online): Peak energies of the cw PL at $T$ = 7 K plotted as function of an externally applied magnetic field in the Faraday geometry. Triangles: sample 1; circles: sample 3

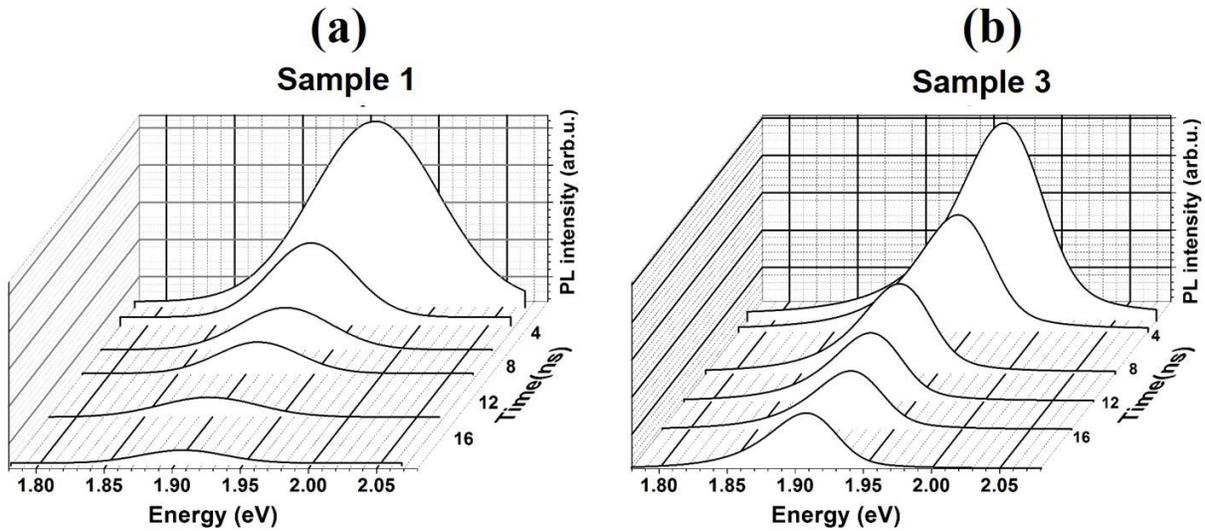

Fig. 5 (color online): Time resolved photoluminescence spectra in cascade form recorded at $B$ = 0, $T$ = 7 K for time delays $\Delta t$ between 0 and 20 ns. (a) Sample 1; (b) Sample 3

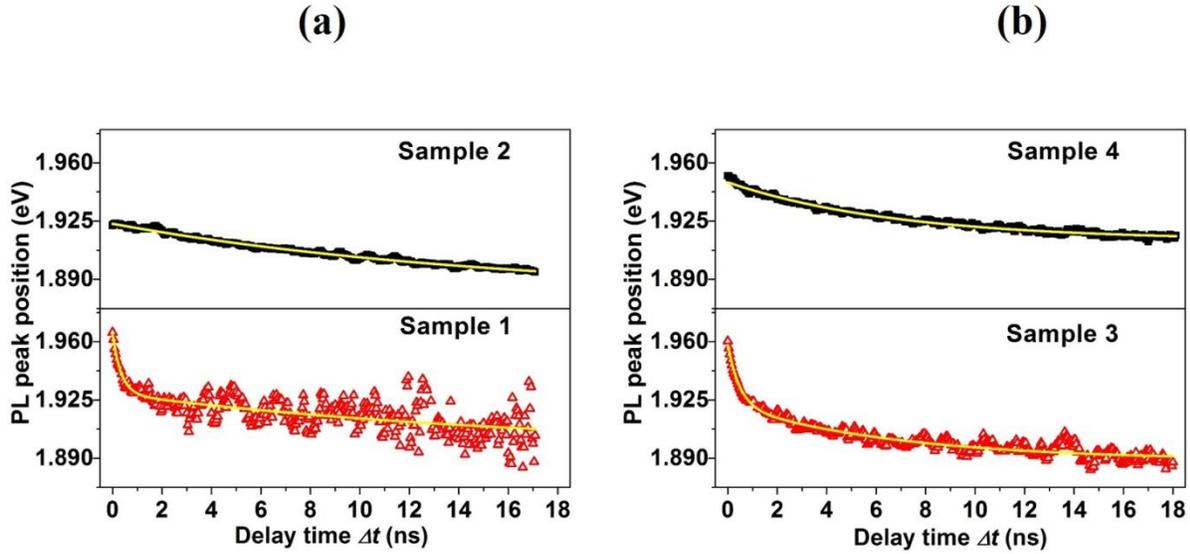

Fig. 6 (color online): Peak energy from TR-PL spectra plotted as function of delay time $\Delta t$ at $T = 7$ K. (a) Upper panel: Non-magnetic sample 2; lower panel: Magnetic sample 1 (b) Upper panel: Non-magnetic sample 4; lower panel: Magnetic sample 3. The solid yellow lines are exponential fits to the data discussed in the text.

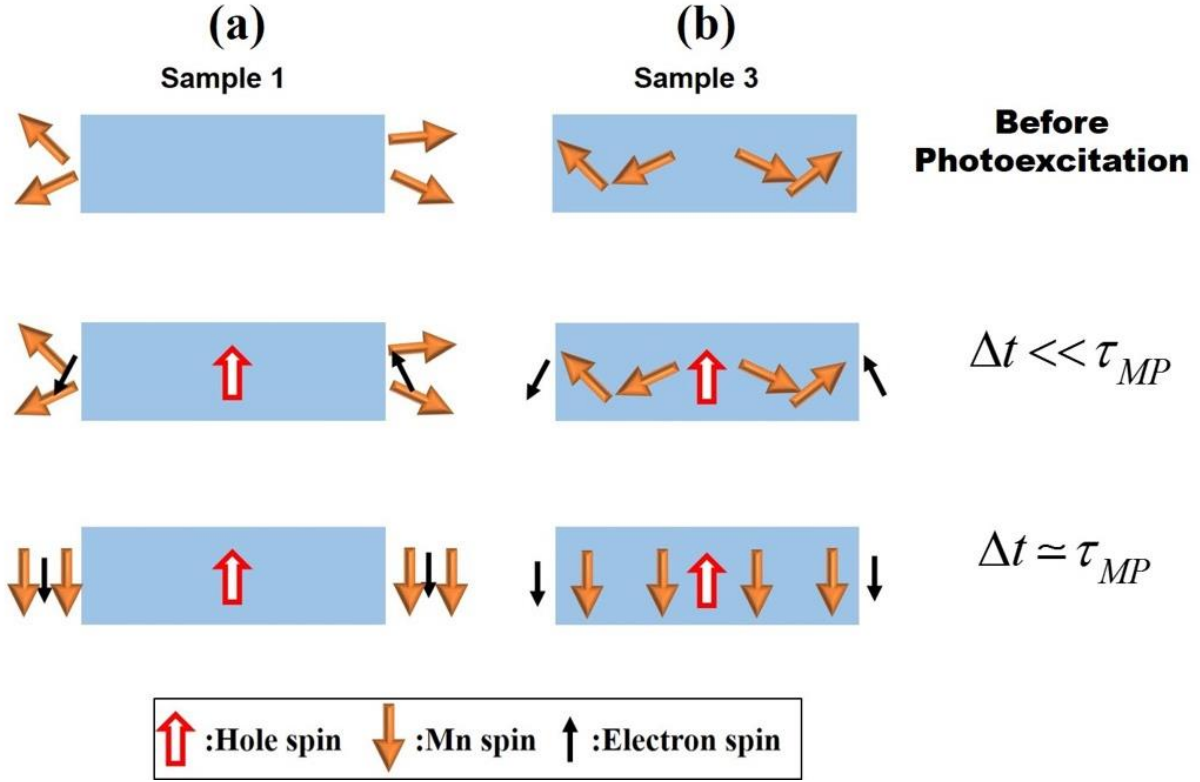

Fig. 7 (color online): Schematic diagram of magnetic polaron formation in: (a) ZnTe/(Zn,Mn)Se sample 1 (b) (Zn,Mn)Te/ZnSe sample 3. The red (black) arrows indicate the hole (electron) spin; orange arrows are used for the manganese ion spins. The blue boxes represent the ZnTe and (Zn,Mn)Te QDs, respectively.

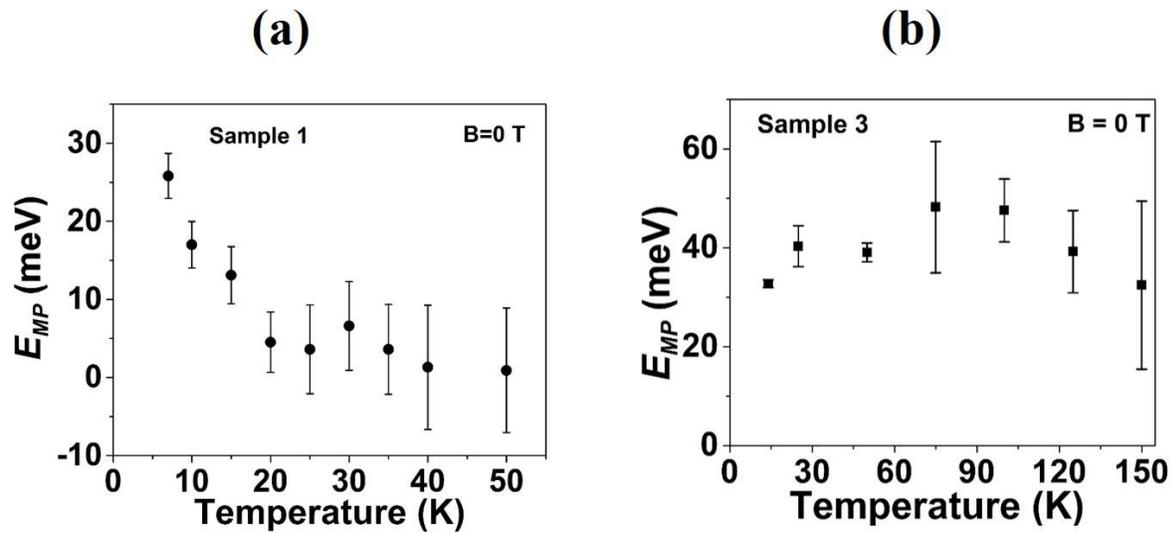

Fig. 8 (color online): Magnetic polaron energy plotted as function of temperature at *B* = 0

(a) Circles: Sample 1 (b) Squares: Sample 3

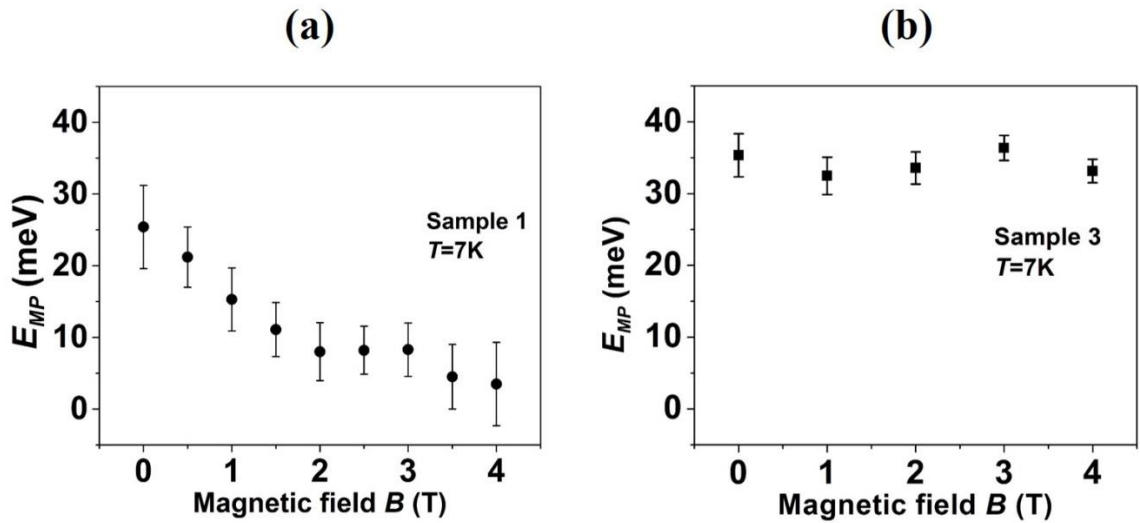

Fig. 9 (color online): Magnetic polaron energy plotted as function of magnetic field at *T*= 7 K

(a) Circles: Sample 1 (b) Squares: Sample 3

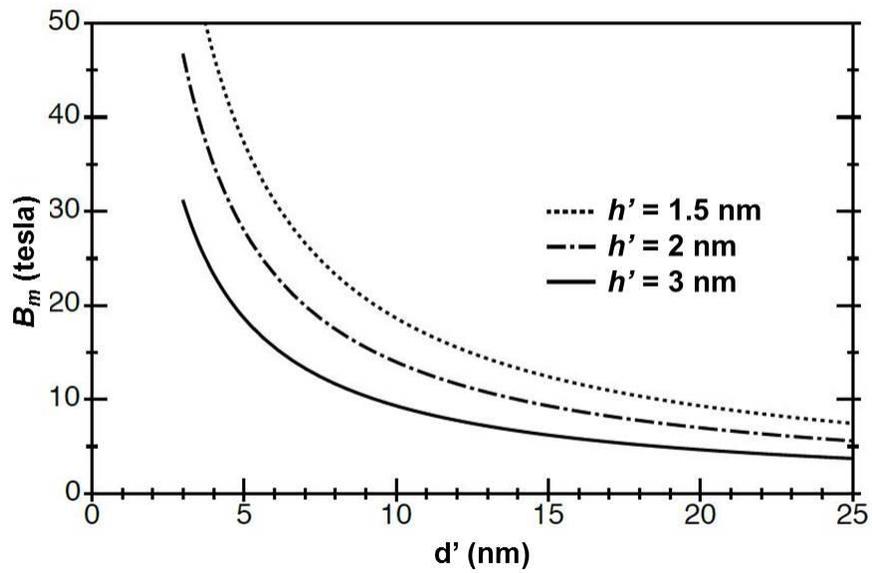

Fig. 10: Molecular field $B_m$ as function of the effective QD diameter $d'$ with the effective QD height $h' = 1.5, 2,$ and $3$ nm.

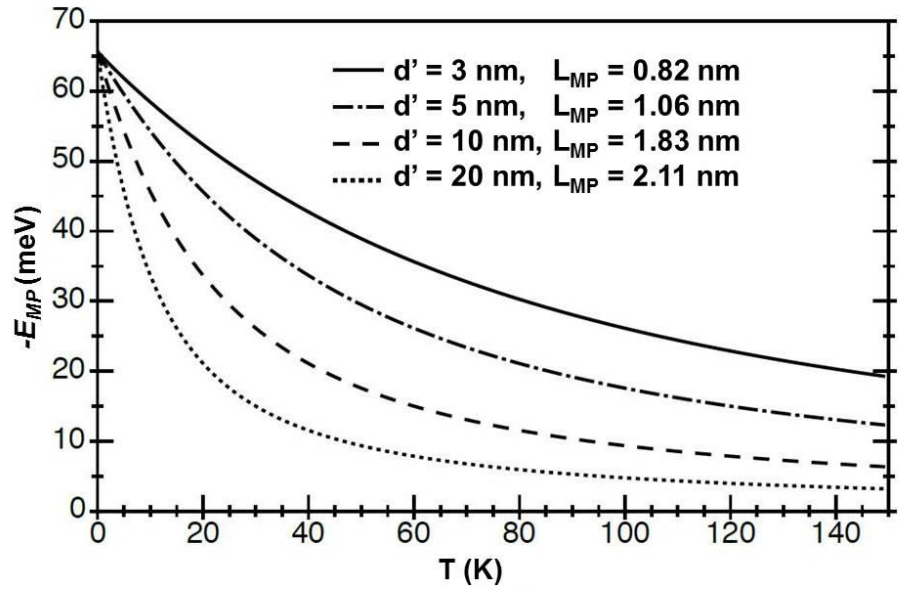

Fig. 11: Average magnetic polaron energy as function of temperature with the QD diameter $d' = 3, 5, 10,$ and $20$ nm.

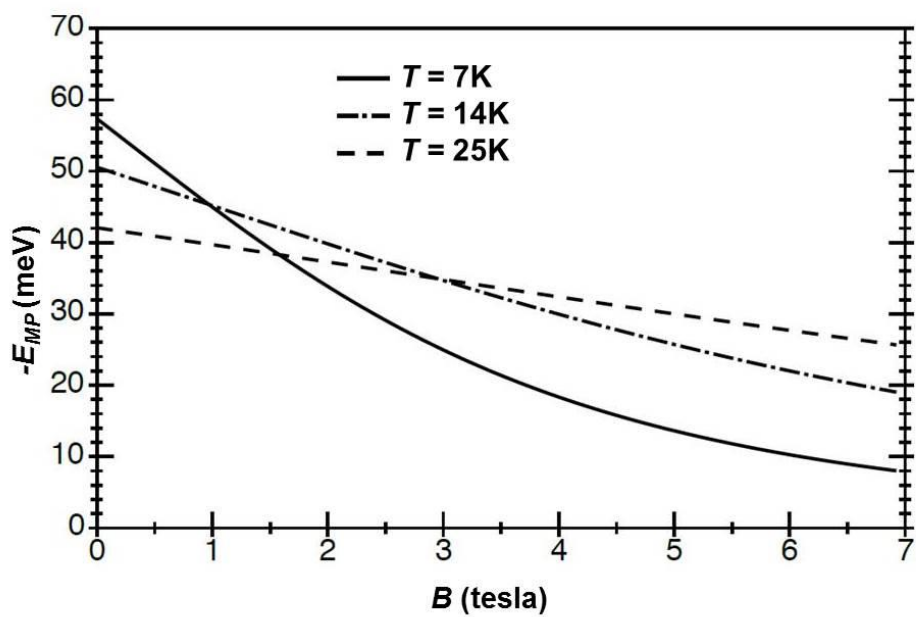

Fig. 12: Average magnetic polaron energy as function of applied magnetic field at $T = 7$, 14, and 25K.